\documentclass[letterpaper]{article}
\usepackage{aaai}
\usepackage{times}
\usepackage{helvet}
\usepackage{courier}
\usepackage{graphicx}
\usepackage{caption}
\usepackage{subcaption}
\usepackage{array}
\usepackage{tabularx}
\begin{document}
%
\title{Could you become more credible by being White? Assessing Impact of Race on Credibility with Deepfakes}

\author{Melissa Welsh, Dillanie Sumanthiran, Ji-ze jang, Md. Rafayet Ali and Ehsan Hoque}
\author{Kurtis Haut, Caleb Wohn, Victor Antony, Aidan Goldfarb}
\author{Ehsan Hoque}
\author{Kurtis Haut, Caleb Wohn, Victor Antony,  \\
Aidan Goldfarb, Melissa Welsh, Dillanie Sumanthiran, Ji-ze Jang \\
Md. Rafayet Ali and Ehsan Hoque \\
University of Rochester\\
Deptment of Computer Science\\
}

\maketitle


\begin{abstract}
 Computer mediated conversations (e.g., videoconferencing) is now the new mainstream media. How would credibility be impacted if one could change their race on the fly in these environments?  We propose an approach using Deepfakes and a supporting GAN architecture to isolate visual features and  alter racial perception. We then crowd-sourced over 800 survey responses to measure how credibility was influenced by changing the perceived race. We evaluate the effect of showing a still image of a Black person versus a still image of a White person using the same audio clip for each survey. We also test the effect of showing either an original video or an altered video where the appearance of the person in the original video is modified to appear more White. We measure credibility as the percent of participant responses who believed the speaker was telling the truth. We found that changing the race of a person in a static image has negligible impact on credibility. However, the same manipulation of race on a video increases credibility significantly (61\% to 73\% with p $<$ 0.05). Furthermore, a VADER sentiment analysis over the free response survey questions reveals that more positive sentiment is used to justify the credibility of a White individual in a video. 
\end{abstract}



\section{Introduction}
\begin{figure*} [t]
    \centering
    \includegraphics{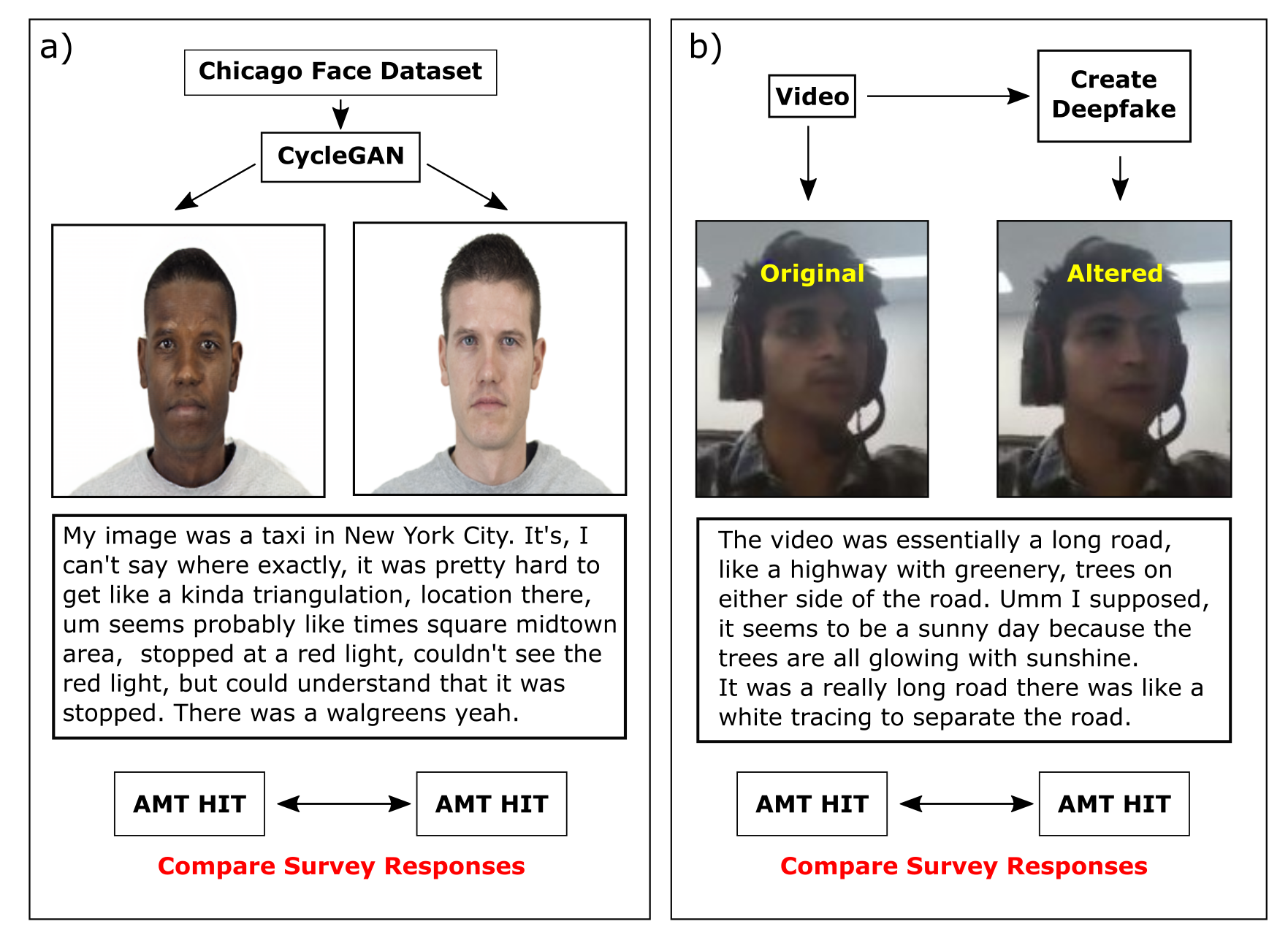}
    \caption{a) Depicts the image condition. We trained a CycleGAN using Chicago Face Dataset to generate unambiguous racial images of a Black and White man. We launched two surveys on Amazon Mechanical Turk (denoted by AMT HIT) and compared the responses. Both surveys have the same audio of the transcript shown. b) depicts the video condition. We took an original video and made subtle racial changes to the face area using deepfacelab. We launched two surveys as in a) and compare the responses. The transcript of what the speaker said is shown and both videos have identical audio.}
    \label{fig:main_fig}
\end{figure*}

Manipulating race to measure its impact has been extensively studied in the past. It has been shown that having a Black sounding name puts you at a disadvantage for employment opportunities compared to having a White sounding name on identical resumes \cite{bertrand2004emily}. Avatars offer a high degree of predictive customization and have been utilized to research impacts of race in virtual reality \cite{groom2009influence} and creating racial empathy that extends to the real world \cite{peck2013putting}. Although names and avatars are both effective ways to create racial perceptions, in the domain of credibility, it remains unclear whether the same findings may extend to subtle manipulation of images and videos. The recent advancements of AI now allows precise manipulation of skin tone, hair type, eye color, and facial features. This opens up an exciting opportunity to probe for race and measure its impact in a way that wasn't possible before. In this paper, we evaluate racial perceptions by leveraging  Deepfakes and a supporting GAN architecture to manipulate facial features and skin color while holding other variables such as accent and clothing constant. This allows us to directly measure the effect of those racial changes. We apply our techniques on a naturalistic dataset on deception collected using the ADDR framework where the participants engage in an activity on a video call \cite{sen2018automated}. Given that the dataset is made of clips from actual video calls recorded by the webcam, we hope that our findings can provide insights for real world videoconferencing situations such as sales, negotiations, job interviews, telemedicine and virtual trials. Virtual trials have become especially important during the COVID-19 pandemic as courts in the US in all 50 states have begun to hold online trials using video conferencing \cite{iowa2020courts,perkins2020courts}.  Virtual testimony is perceived to be less credible that in-person testimony. \cite{landstrom2010court,landstrom2005witnesses,walsh22effective}. There could be many reasons for this, but understanding whether perception of race is contributing to this phenomenon may help future court proceedings. Even for low stakes video calls, credibility assessments of our conversational partners impact communication. Therefore, understanding how racial perceptions influence credibility in these domains matters. Perception of race is a complex issue composed of variables such as class, education, accent, skin tone, etc. In this paper, we look at racial perceptions through the lens of modifying visual representation (skin tone and facial features).

We designed an experiment consisting of 4 surveys (see fig. \ref{fig:main_fig} to begin the process of answering these research questions:
\begin{itemize}
    \item How do racial perceptions in videos and images influence an individual's credibility?
    \item What public sentiments are associated with the credibility of a perceived race?
\end{itemize}

 To assess credibility, we employ Amazon Mechanical Turk (AMT) to crowd-source responses for our surveys. An online worker for AMT is called a mTurker who accepts a Human Information Task (HIT) and is paid for completing the human labor associated with that HIT. In our case, a mTurker watches a video or listens to an audio with a still image, then completes a survey. In our surveys, we take credibility to mean the percent of participants who believed that the speaker in the video or audio was telling the truth. In total, we have 4 surveys with 800 total responses from the participants (For demographic information see fig. 2). The first 2 surveys test the condition of what happens when a different still image is shown to the participant while the same audio file is played, i.e., for one survey an image of a Black person is shown to the participant while in the other survey, an image of a White person is shown to the participant (see fig. 1a). The next 2 surveys test the condition of showing a different video to the participants. In one survey, a video is shown.  In the other survey, we show an altered version of the same video where the only difference is we make the speaker appear White (see Fig.1b). We used the CycleGAN \cite{zhu2017unpaired} to generate image mappings of White to Black and vice versa to test the image condition. For the video condition, we used an encoder decoder architecture to perform a face swap; essentially changing the visual representation of a South Asian person to appear more White (in terms of skin tone, facial features) using DeepFaceLab \cite{deepfacelab}. For each of the four surveys, 200 participants (total 800) responded. We then performed statistical analysis on the survey responses to determine if the credibility differences for each condition were meaningful. 



In summary, we find and recommend the following:
\begin{itemize}
    \item Those who believed the person was White in the altered video were more likely to say the person was telling the truth. 
    \item There is no statistical difference in credibility between the image conditions when participants listened to the same audio.
    \item Participants described the altered video using words with \textit{more positive} sentiment in the free response survey questions.
    \item Subtle modifications to alter racial perception may impact credibility in video recordings from prior videoconferencing sessions.
\end{itemize}

Although creating a high quality Deepfake in real-time during a videoconferencing session is not available right now, in the near future this will be possible. Regardless of the situation surrounding the video call, it is likely that people may have the ability to change their perceived race (users can already change their gender or skin complexion on Snapchat). Our findings indicate that this action could have an effect on the credibility of that person. We make the following recommendation to video-conferencing or social media companies, should they decide to enable this feature in the future:
\begin{itemize}
    \item Complete transparency and full disclosure of the race change to all parties on the video call. (e.g., they must be notified and there should be an identifying mark on the window of that persons video during the call).
\end{itemize}

\section{Background}
\subsection{Racial Biases within Societies}
Prior studies within the context of racial discrimination aim to tackle the problem solely through using sociological metrics of quantification. Perception of racial identity has been observed to depend on various aspects of race, including skin color \cite{maddox2002cognitive} and accents \cite{dixon2002accents}. When assessing racial discrimination in context of skin color, prior works illustrate how people of different races tend to distrust each other the most \cite{smith2010race}. The level of racial discrimination can even be broken down further into the skin color gradient \cite{leonardo2004color}. For example, even when race is held as a constant and only skin color fluctuates, dark-skinned Blacks were 11 times more likely to experience racial discrimination than their light-skinned counterparts \cite{klonoff2000skin}. These studies suggest that skin color alone may cause a person to be discriminated against. 
In addition to skin color and race, the accent of a person can also have a significant influence on how others perceive them. This is often found in incidents involving people who speak African American Vernacular English (AAVE). There have been many court cases in which the accent of African Americans negatively influence the outcome of a trial \cite{rickford2016language}. 
Moreover, people with a non-native accent may cause listeners to perceive them as being less credible because their accents serve as a signal or because listeners find their speech to be difficult to process \cite{lev2010don}. Through their work, Shiri Lev-Ari and Boaz Keysar showed that participants in their study perceived trivia statements such as “Ants don't sleep” to be true less frequently when spoken by a non-native English speaker than a native speaker. This lends credence to the idea that an individual’s socio-political background may have a significant impact on how credible they are perceived to be. Similarly, Günaydin et al demonstrated that objective facial resemblance to a significant other influences snap judgments of liking automatically, effortlessly, and without conscious awareness \cite{GUNAYDIN2012350}. This yields further evidence to the possibility that race may impact credibility given the large proportions of individuals who grow up in communities which are segregated or dominated by one racial group.

However, there is very limited information in relevant literature assessing the impacts of perceived race on credibility through modifying visual representation exclusively. This could be due to the fact that one can not easily change their physical features of race nor can they replicate life experience to measure the effect of those changes. However, when we communicate online with images or videos, these constraints do not exist. Because we can modify visual representation algorithmicly on the computer, we can research the effect this has on credibility when communicating face to face over digital media.  

\subsection{Employing AI for Investigating Racial Bias}
In order to more objectively assess how skin color and facial features effect one’s credibility, techniques such as Cycle GANs and Deepfakes are the most suitable ones. Those techniques can keep features such as accent, clothing, gestures, and facial expressions constant. Cycle GANs have previously been used within research applications to make image processing training sets more inclusive to various skin tones. In the work of “Fairness GAN”, Sattigeri illustrates the CycleGAN’s power in constructing an extension to the Celebfaces Attributes to be demographically inclusive, one portion being skintone\cite{sattigeri2018fairness}. Other work with Cycle GAN looks at applications such as artificial makeup, which can vary in degree of heaviness when applying synthetic makeup \cite{chen2019beautyglow}.  Deepfakes also hold significant value in manipulating skin tone and facial features of videos. Although researchers have worked on creating algorithms to detect  computer manipulated images and videos \cite{korshunov2018deepfakes}, and humans are likely to be more accurate at this task at present\cite{korshunov2020deepfake}, the future is anything but certain. These algorithms continue to improve, generating more realistic images at a startling pace. In 2019, Karras et al introduced StyleGAN which demonstrated an innovative architecture able to generate more realistic images \cite{karras2019style}. In this paper, we leverage Deepfakes and Cycle-GANs to create subtle racial changes to help understand the nuances of race perception and credibility in video call environments.

\section{Methods}

The credibility and associated language sentiment for different racial perceptions was measured using responses from 800 participants recruited from Amazon Mechanical Turk. Generally, mTurkers form a more diverse population than the standard internet population or the populations of college students typically used in laboratory experiments \cite{buhrmester2016amazon}\cite{horton2011online}\cite{paolacci2010running}.  These participants were given one of the four surveys which were designed to measure two conditions; image and video.  Among the 800 participants 305 gave us their demographic information (shown in Fig. 2a, 2b and 2c). These participants were compensated 2 dollars for their time completing the surveys.

\begin{figure}[tbh!]
     \centering
     \begin{subfigure}[]{.4\textwidth}
         \centering
         \includegraphics[width=\textwidth]{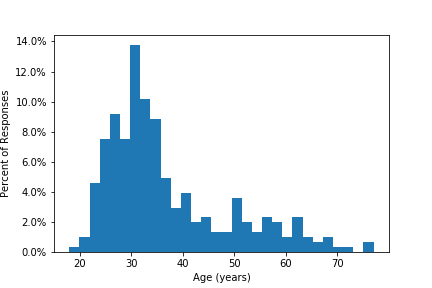}
         \caption{Age Distribution}
         \label{fig:turk_age}
     \end{subfigure}
     \begin{subfigure}[]{.35\textwidth}
         \centering
         \includegraphics[width=\textwidth]{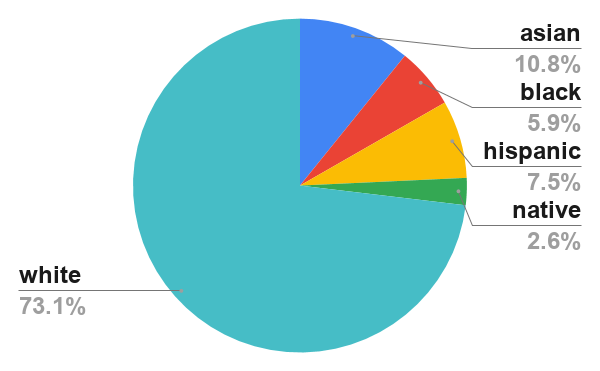}
         \caption{Race}
         \label{fig:turk_race}
     \end{subfigure}
     \begin{subfigure}[]{.35\textwidth}
         \centering
         \includegraphics[width=\textwidth]{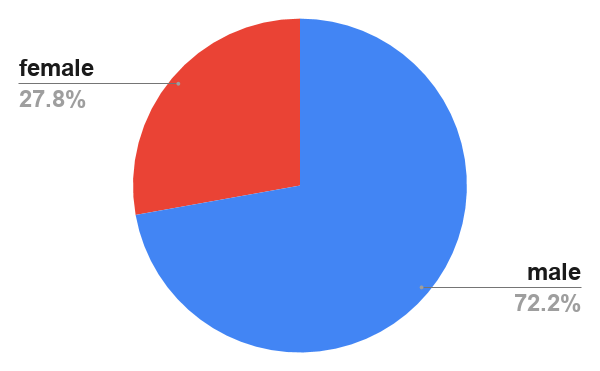}
         \caption{Gender Identity}
         \label{fig:turk_gender}
     \end{subfigure}
     \\
     \begin{subfigure}[]{.35\textwidth}
    	\centering
    	\begin{tabular}{|c|c|c|}
    		\hline
    		Country & Count & Percent \\
    		\hline
    		US & 263 & 86.23\% \\
    		Brazil & 16 & 5.25\% \\
    		India & 6 & 1.97\% \\
    		Italy & 4 & 1.31\% \\
    		Bangladesh & 2 & 0.66\% \\
    		China & 2 & 0.66\% \\
    		Ireland & 2 & 0.66\% \\
    		Spain & 2 & 0.66\% \\
    		U.K. & 2 & 0.66\% \\
    		Argentina & 1 & 0.33\% \\
    		Bahamas & 1 & 0.33\% \\
    		Canada & 1 & 0.33\% \\
    		Hong Kong & 1 & 0.33\% \\
    		Indonesia & 1 & 0.33\% \\
    		Mexico & 1 & 0.33\% \\
    		\hline
    	\end{tabular}
    	\caption{National Origin}
    	\label{fig:turk_nations}
    \end{subfigure}
        \caption{Self-reported demographic information of 305 participants}
        \label{fig:three graphs}
\end{figure}

\subsection{Design of Experiment}
\noindent \textbf{Credibility Ground Truths} \newline
We used one audio clip and one video clip from the publicly available UR Lying dataset, collected using the ADDR framework \cite{sen2018automated}.  Each clip is 30 seconds in length and encompasses the speaker answering a question \textit{What was your image?}. The speakers in the video are shown an image prior to asking this question. The ADDR framework instructed the speaker to lie or tell the truth about their image. In this particular instance, the speakers in both image condition and in video condition described their image as it is (i.e., telling the truth). From the audio and video recordings, we designed four separate surveys to test two conditions. Two surveys tested an image condition and two surveys tested a video condition. We gathered a total of 800 responses from the four surveys each survey having 200 participants.  

\subsubsection{The Image Condition}
We were interested to see how still images of people from different races paired with the same audio can influence credibility. The still image was designed to elicit a specific racial perception. We were interested in comparing credibility for a White person versus a Black person. Instead of using a real image from the dataset, which introduces noise such as clothing style and features unique to an individual of that race, we generated our own generic race representations to produce racial mappings of Black to White and vice versa as shown in fig. 1a. We trained a CycleGAN using 4000 images from the Chicago Face Dataset (CFD) which learned how to conceivably change complex racial features such as eye color, lip shape, hair type and skin color \cite{CycleGAN2017}. The participants either saw the White image in the survey or the Black image in the other survey, while listening to the same audio. We then compared the responses from these two surveys to see if the still image shown to the participants influenced the speakers credibility or attributed language sentiment. 


\subsubsection{Video Condition}
While the image condition created clear, unambiguous representations of race, the video condition explores the nuances of racial perception by making subtle modifications to the facial region. 
In this condition, participants either watched the original video in one survey or an altered video in the other survey. In the altered video, the person from the original video is made to appear more White using a DeepFaceLab \cite{deepfacelab} face mapping. We do it by taking the face of a White participant from the UR Lying database and map it onto the face of a non white participant as seen in fig 1b. We then create 2 surveys where one survey shows the participant the original video and the other shows the altered video. We then compare the responses on the credibility and language sentiment of the speaker. 



\subsubsection{Survey}
To measure the effect of an individual's perceived race on that individual's credibility we created four independent surveys. Two of the surveys played an audio file from the UR Lying Database accompanied with either a Black or White still image. The other two surveys played either an original video from the UR Lying database who is South Asian or an altered video where the person is made to appear more White. For each survey, we collect information on whether the participant thought the speaker was lying or telling the truth and ask the participant to justify his/her answer using a free response. The questions we asked the participants remain the same in each survey and can be found in table 1. All together, we collect 200 responses per survey totaling 800 responses.

\begin{table}[!htb]
  \caption{Questions for the surveys}
  \label{tab:mturk_survey}\centering
  \begin{tabularx}{3.3in}{X}
    \hline
    \textbf{Questions} \\
    \hline
    Please briefly (1 or 2 sentences) describe what the person in the video said. \\
    \hline
    Do you think the speaker was telling the truth?\\
    \hline 
    How confident are you about the whether the speaker was lying/telling the truth? (10 means you are certain, 1 means you have no idea) \\
    \hline
    What made you think the person was lying or telling the truth? \\
    \hline
    Use a few adjectives to describe the characteristics of the speaker \\
    \hline 
    What race do you think the speaker is? \\
    \hline
    What do you think the socioeconomic status of the speaker is? \\
    \hline
    Do you think the video was authentic?\\
    \hline
    What is your age? \\
    \hline
    What do you consider to be your gender? \\
    \hline
    Where did you grow up? (i.e. where did you spend most of your first 12 years?) \\
    \hline
    What do you consider to be your socioeconomic class? \\
    \hline
    What do you consider to be your race? \\
    \hline
  \end{tabularx}
  \vspace{-3.5mm}
\end{table}

\subsection{Analysis}
\subsubsection{Pre-processing to Identify Perceived Race}
We included a question asking the participants to identify the race of the speaker with the goal of detecting the perceived race. Since our research question concerns how perceptions of race influence credibility, we had to filter out responses which did not "correctly" identify the race (i.e., "Black" in the case of the audio with a still image of a Black man, "White" in the case of the audio with a still image of a White man, "Asian" in the case of the original video of a South Asian speaker, and "White" in the case of the altered video). Thus, after the pre-processing step is complete in our analysis pipeline, we are left only with the responses who perceived the intended race for each survey.

\subsubsection{Assessing perceived credibility} 
We asked the participants whether or not they thought the speaker was telling the truth in each survey in a binary fashion (i.e., yes or no). For the image condition, we analyzed the responses from the two surveys where this image is either Black or White as described earlier. We took the number of respondents who thought the speaker was telling the truth in both cases and compare them against each other using a proportions Z-test. Similarly, for the video condition, we analyze the responses from the two surveys where the video is either the original or altered video as described earlier. Following the same routine for the image condition, we take the number who thought the speaker was telling the truth in both surveys and compare them using a proportions Z-test.

\subsubsection{Analyzing Sentiment} 
We also asked the participants to give their rationale for why they thought the speaker was lying or telling the truth as well as to give a few adjectives to describe the characteristics of the speaker in a free response form. We analyzed these text-field questions using a VADER sentiment analysis \cite{gilbert2014vader}. VADER uses simple rules and a lexicon built by averaging sentiment analysis done by participants from Amazon Mechanical Turk. Its heuristic rules were designed based on statistical analysis of tweets, and it was tested against other models across multiple domains using AMT to generate a ``wisdom of the crowd" ground truth. A major reason why we chose VADER is because VADER was created and tested using AMT surveys (a population similar to ours). We first perform a compound sentiment analysis on the text of each response, which outputs a number between -1 and +1 for that response. Thus for each survey, we now have an array of values between -1 and +1. We take the arrays from the two surveys associated with the image condition and compare them against each other using a Mann Whitney U test. Likewise for the video condition, we take the arrays from the two surveys for that condition and compare them against each other. This allows us to evaluate whether more positive sentiment is associated with a specific racial perception. We do the same sentiment analysis over the responses for the question which asked the participants to use a few adjectives to describe the characteristics of the speaker.

\section{Results}
Fig. 3a, 3b, 3c and 3d show how the pre-processing was done on the responses for the analysis. Because we are interested in how racial perceptions influence credibility over digital media, we analyze the responses who believed the intended racial perception as described in the pre-processing section.

\begin{figure}
     \centering
     \begin{subfigure}[h!]{.4\textwidth}
         \centering
         \includegraphics[width=\textwidth]{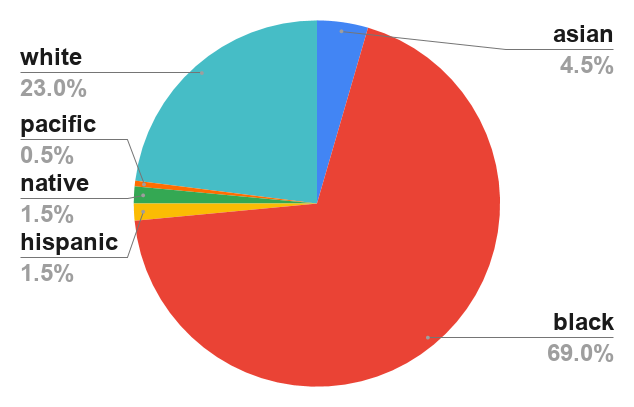}
         \caption{Participants Race Responses for Black Image}
         \label{fig:turk_age}
     \end{subfigure}
     \hfill
     \begin{subfigure}[h!]{.4\textwidth}
         \centering
         \includegraphics[width=\textwidth]{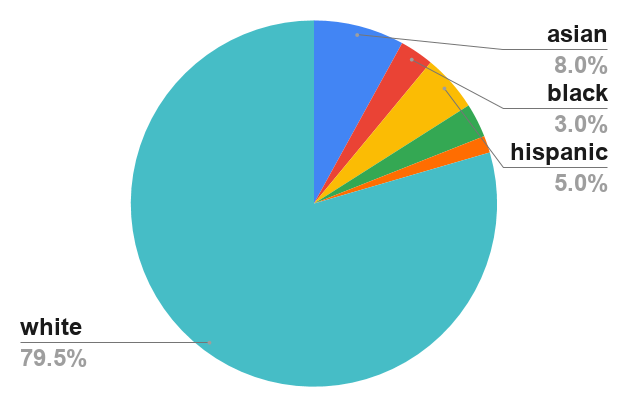}
         \caption{Participants Race Responses for White Image}
         \label{fig:turk_race}
     \end{subfigure}
     \hfill
     \begin{subfigure}[h!]{.4\textwidth}
         \centering
         \includegraphics[width=\textwidth]{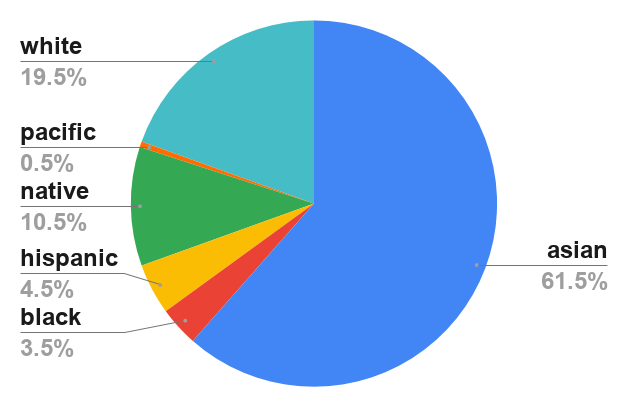}
         \caption{Participants Race Responses for Original Video}
         \label{fig:turk_gender}
     \end{subfigure}
     \begin{subfigure}[h!]{.4\textwidth}
         \centering
         \includegraphics[width=\textwidth]{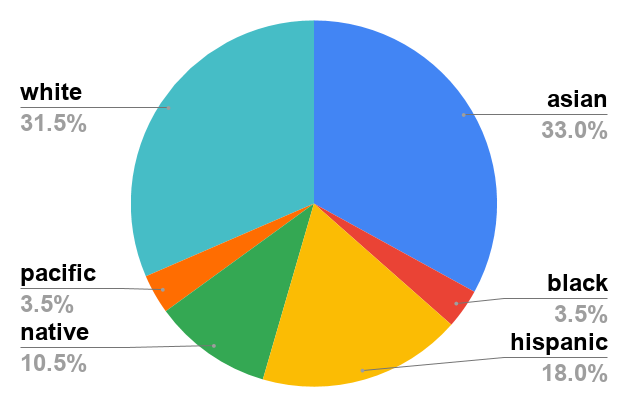}
         \caption{Participants Race Responses for Altered Video}
         \label{fig:turk_gender}
     \end{subfigure}
        \caption{Results for the racial perceptions of the participants from the pre-proccessing question \textit{``What race do you think the speaker is?"}}
        \label{fig:race-pies}
\end{figure}

\subsection{Credibility Analysis}
We take credibility to mean the percent of respondents who said that the speaker was telling the truth. In the image condition, 72.3\% believed the speaker was telling the truth when an image of a White person was shown vs 70.3\% when an image of a Black person was shown. There was no statistical significant difference in this condition. But in the video condition, 61.0\% of participants believed the speaker was telling the truth after watching the original video of the darker-skinned South Asian speaker, and 73.0\% of participants believed the speaker was telling the truth after watching the altered video featuring the White speaker. This difference was significant ($p<.05$). Table \ref{tab:credibility_results} shows these results. 

\begin{table}[!htb]
  \caption{Summary of Results. ``n" is the number of responses after pre-processing. ``Credibility" is the percent of responses that believed the speaker was telling the truth. * indicates p $<$ .05.}
  \label{tab:credibility_results}\centering
  \begin{tabularx}{3.3in}{c|c|c}
    \hline
    \textbf{Survey} & \textbf{n} & \textbf{Credibility} \\ 
         \hline
         Black Image & 138 & 70.29\% \\
         White Image & 159 & 72.33\%  \\
         \hline
         Original Video (South Asian) & 123 & *60.98\% \\
         Altered Video (White) & 63 & 73.02\%  \\
        \hline 
  \end{tabularx}
  \vspace{-3.5mm}
\end{table}

\subsection{Sentiment Analysis}

\begin{figure}[!hbt]
    \centering
    \subfloat{\includegraphics[width=3.3in]{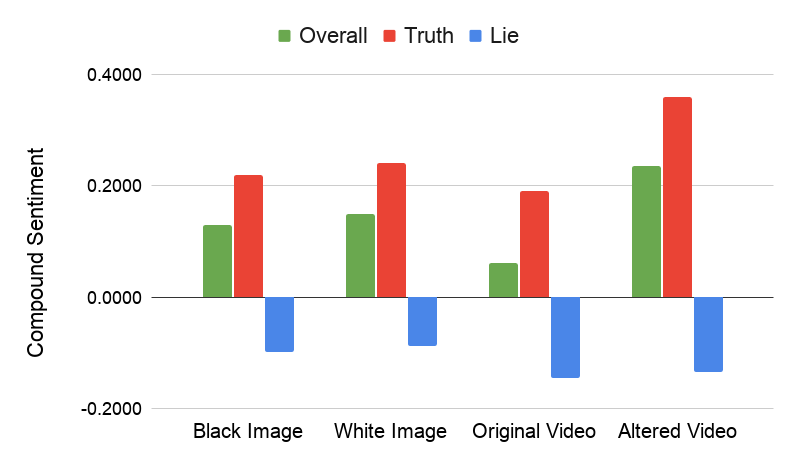}}
    \\
    \subfloat{
      \begin{tabularx}{3.3in}{c|c}
      \hline
        \textbf{Survey Comparison} & \textbf{VADER Sentiment} \\
        \hline
            Black Image Overall & 0.13  \\
            White Image Overall & 0.149  \\
            \hline
            Black Image Truth & 0.22 \\
            White Image Truth & 0.24  \\
            \hline
            Black Image Lie & -0.098  \\
            White Image Lie & -0.088 \\
            \hline 
            Original Video Overall & **0.061  \\
            Altered Video Overall & 0.236  \\
            \hline
            Original Video Truth & **0.19  \\
            Altered Video Truth & 0.359 \\
            \hline 
            Original Video Lie & -0.145  \\
            Altered Video Lie & -0.133  \\
            \hline 
      \end{tabularx}
    }
    \caption{Results from VADER compound sentiment analysis of responses to the question \textit{``What made you think the person was lying or telling the truth?"}.  Overall indicates the mean of the compound sentiment regardless of whether the participants believed the speaker or not. `Truth' indicates those respondents who thought the speaker was telling the truth and `Lie' indicates those participants who thought the speaker was lying. *Indicates p $<$ 0.05 and ** when p $<$ 0.01.}
    \label{fig:vader_why}
    \vspace{-3.5mm}
\end{figure}

\begin{figure}[!hbt]
    \centering
    \subfloat{\includegraphics[width=3.3in]{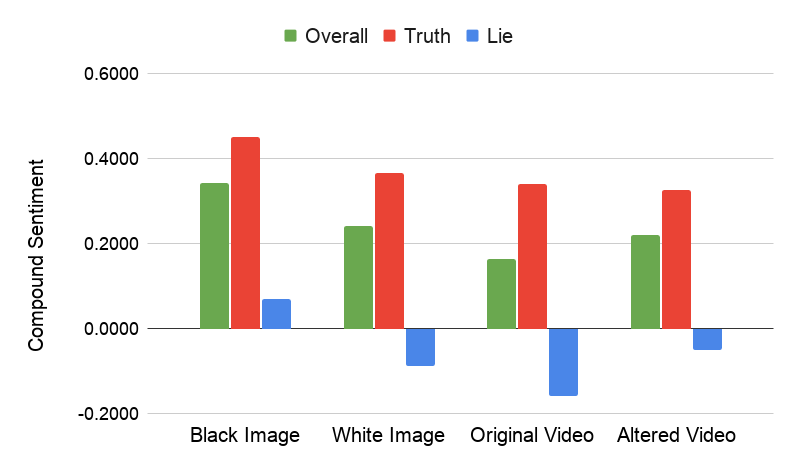}}
    \\
    \subfloat{
      \begin{tabularx}{3.3in}{l|r}
      \hline
        \textbf{Survey Comparison} & \textbf{VADER Sentiment} \\
            \hline
            Black Image Overall & *0.342  \\
            White Image Overall & 0.241 \\
            \hline
            Black Image Truth & *0.451  \\
            White Image Truth & 0.367  \\
            \hline
            Black Image Lie & *0.07  \\
            White Image Lie & -0.088 \\
            \hline 
            Original Video Overall & 0.163  \\
            Altered Video Overall & 0.221  \\
            \hline
            Original Video Truth & 0.34  \\
            Altered Video Truth & 0.325 \\
            \hline 
            Original Video Lie & -0.159  \\
            Altered Video Lie & -0.05  \\
            \hline 
      \end{tabularx}
    }
    \caption{Results from VADER compound sentiment analysis of responses to the question ``Use a few adjectives to describe the characteristics of the speaker?". Overall indicates the mean of the compound sentiment regardless of whether the participants believed the speaker or not. `Truth' indicates those respondents who thought the speaker was telling the truth and `Lie' indicates those participants who thought the speaker was lying.  *Indicates p $<$ 0.05 and ** when p $<$ 0.01.}
    \label{fig:vader_adj}
    \vspace{-3.5mm}
\end{figure}

We analyzed the responses to the open-ended questions, \textit{``What made you think the person was lying or telling the truth?''}  and \textit{``Use a few adjectives to describe the characteristics of the speaker''} . We used the compound sentiment score from VADER, which measures sentiment on a scale from -1 to +1, with +1 indicating the text is very positive. The results are contained in Figures \ref{fig:vader_why} and \ref{fig:vader_adj}.

In all text responses used to justify their credibility decisions, respondents who thought the speaker was telling the truth had significantly higher sentiment than those who thought the speaker was lying. We found that the altered video had higher sentiment overall compared to the original video (0.236 vs. 0.061, p $<$ 0.01) . In particular, respondents were more positive when talking about a perceived White person telling the truth than a perceived South Asian person telling the truth (0.359 vs 0.19, p $<$ 0.01). When respondents were justifying dishonest speakers there was no significant difference in sentiment for the videos.

In the text responses containing the adjectives used to describe the speaker, respondents used more positive sentiment to describe the Black image compared to the White image overall (0.341 vs 0.241 p $<$ 0.05).  This difference remains significant regardless of whether the participants thought the speaker was telling the truth.

\section{Discussion}
\subsection{Nuances of Racial Perception Could Reveal Unconscious Bias}
Our decision to use still images of a White person versus a Black person was done to make the perception of race less ambiguous. This can be seen by looking at the distribution of survey responses in figure \ref{fig:race-pies}. It may seem like making the differentiation of race more clear would enable measuring the effect on credibility more accurate. However, when this racial difference is accentuated, it is also possible that the survey participants compensate their credibility evaluations based on their awareness of what is being measured due to the Hawthorne effect \cite{wickstrom2000hawthorne}.

For this reason, we also choose to do a more nuanced racial modification with the video condition. In this condition, the racial perception is much more ambiguous. This can be seen by the distribution of the survey responses in figure \ref{fig:race-pies}. We speculated, that perhaps if the racial changes were subtle enough they could even go unnoticed. This could allow unconscious bias to influence the survey responses accessing credibility.
It remains part of our future work to probe and compare different racial filters including a bi-racial White and Black speaker for the video condition. More work is needed to also explore the bi-directional alternation of races to understand the whole spectrum of racial perception.

\subsection{Image Condition Versus Video Condition Credibility Comparison}
We did not observe a difference in credibility in the still image condition compared to the video condition. Our original hypothesis was changing the racial perception of the speaker to be White would increase their credibility. Although we see some evidence of that effect with a 70.3\% credibility rating with the Black still image shown compared to a 72.3\% credibility rating with the still White image shown, this difference is not significant. 

Yet with the video condition, we see that making the speaker appear White does have a tangible influence on their credibility rating (61.0\% increased to 73.0\% when the participants perceived the speaker as White). This provides some evidence that appearing more White may increase credibility in videoconferencing platforms. However, that conclusion does not explain why we see an increase in credibility for the video condition but not for the image condition. It is possible that for the image condition with two racially pronounced pictures, the respondents were more aware of the racial bias and didn't allow race to influence their judgement. However, the video condition included a more nuanced alteration along with nonverbal dynamics for the respondents to process towards making a decision. As a result, it is possible for race to become a factor towards measuring credibility. 


\subsection{Comparing Feelings and Sentiment Towards Race}
 In order to see if perceived race had an effect on sentiment, we compared the responses for surveys that are associated with each condition (image and video). For each condition, we measured and compared the overall sentiment of the open-ended text responses.  We observed no differences in sentiment for the image condition given by the participants for their justification of speaker credibility. However, for the video condition, again we see statistically significant differences. We found more positive sentiment overall in the words used by participants to justify their decision and that this difference is more pronounced when comparing truthful text responses. Given that we observed higher sentiment scores in the altered video, it was likely due to the participants being under the impression that the speaker was White. Yet as we do not see this mirrored in the image condition, there could be more to the story. 

When we compared the text responses describing the characteristics of the speaker, we found higher sentiment in the Black image survey compared to the White image survey. We found this difference is significant regardless of if the participants believed the speaker was telling the truth or not.  The higher overall sentiment attributed to the speaker perceived to be Black indicates the Hawthorne effect at work due to its contradictory nature. Because the positive sentiment is not reflected in the justification for the decision made regarding the truthfulness of the speaker, nor is the credibility rating higher for the Black speaker, it is possible that the positive adjectives used by the participants are insincere.  Looking at the demographics, a majority of the participants are young, white males (see fig. \ref{fig:three graphs}). This particular age/race group, given the current political climate, may be exercising caution when addressing the Black community specifically. A careless choice of words may be viewed as insensitive or could potentially result in being labeled as a White supremacist or racist. In the case of the mTurker, such a complaint could have their license suspended and cause them to be out of work. As a result, for the image condition, it is possible that the participant is cognizant of what is being measured and therefore adjusts his/her behavior accordingly. However, with the video condition, due to the subtleties of the racial adjustments, perhaps we observe the unconscious bias associated with the participants preferring the speaker when perceived to be White. 

\subsection{Recommendations for Video Conferencing Systems}
In the future, our virtual identities will become increasingly important. In the physical world, attributes like skin tone, hair type, eye color and facial features are  hard to change. These constraints for altering one's physical appearance do not exist when we present ourselves in digital media. If AI can alter how we are perceived, it is important to speculate the effect those changes can have on individual credibility in videoconferencing environments. We argue that altering perceived race without informing the audience involved is unethical. We recommend that videoconferencing systems require full disclosure to all parties on the video call when someone has chosen to alter their race. The system should be the entity to provide this disclosure as holding the individual accountable to disclose that information may be unreliable. We propose this disclosure be given one time to each person who joins the call through a notification system. We also recommend that the videoconferencing system implement complete transparency for the duration of the call. We propose that the system implement a distinguishing mark in the frame of the person who is altering their racial appearance. The mark should be discrete, yet serve as a constant reminder to the audience. Lastly, we recommend that the videoconferencing systems implement an appropriate encryption scheme to identify videos sent and received over their channel. Otherwise, a nefarious entity could manipulate the videos unbeknownst to the system and therefore its users. This action would bypass the intentions of full-disclosure and complete transparency. 


\subsection{Limitations}
\subsubsection{Definitions of ``Race"}
How human beings racially profile each other is a complex phenomenon composed of many factors. Among these factors, linguistic features of speech such as accents is a major component. This paper instead focuses on what happens when the visual representation of an individual changes to appear White or more White. 

\subsubsection{Issues of Video Realism}
A potential confounding factor in our experiments is the believability of the altered video. It is conceivable that if a participant found the video to be contrived/altered, that this could influence their evaluation of the speakers credibility.

To address the issue of video realism as a confounding factor in our study, we posed the question 
\textit{``I found the video to be authentic"} in both the original and the altered video surveys. For the original video, 10\% of participants found the video to be not authentic. Compared to the altered video, 15\% of participants thought the video was not authentic. After running a proportions z test we concluded that there is no difference between these surveys regarding the perceived authenticity of the videos shown to the participants (p $>$ 0.1).

\subsubsection{Collecting Data from Amazon Mechanical Turk}
We use AMT as a mechanism to draw a sample of the general population. We must recognize that this surveying technique will suffer from some degree of population bias. When we look at the demographics of the survey responses, there is a bias towards younger, White males from the US. We could have chosen other surveying options. Unfortunately, the cost of bringing people into the lab to complete the surveys was made burdensome due to the the social distancing stipulation of the pandemic. It also would have been far more expensive and would implement a population bias of a different kind. Considering the trade-offs, AMT was our best option given its experimental validity and ability to crowd-source a large number of responses effectively. 

\subsubsection{Realism of Credibility Ground Truths}
Another issue with our experimental findings is the ground truths for the credibility assessment videos that were shown to the participants. We sacrifice realism in order to guarantee the ground truths that were collected during an experimental setting. The deceptions that occur over face to face, computer mediated platforms may be very different from those obtained during an experimental context, and the influence of racial perceptions could likewise be as well. Despite that acknowledgement, the videos used were of study participants participating in a computer mediated deception experiment. Thus we expect the video to be good to use for exploring credibility in the domain of digital media.

\subsection{Future Work}
We note here that we cannot directly compare the difference of racial perceptions created via images and audio versus video in how they affect credibility over video calls. We would need many examples using consistent image frames and audio to begin to draw a generalized conclusion on that topic. We leave the fine grained analysis of still image versus video for future work as the implications are important to understand. 

Lastly, this paper focuses on examining what happens when the visual representation of race is changed with AI and how that influences one's credibility. In the future, we will isolate specific accents such as African American Vernacular English (AAVE) and compare that against standard American English to observe the direct effect this has on an individual's credibility. AI can alter visual appearance and should also be able to modify accent. This will enable new insights due to the isolation and fine grained control of racial audio features. Understanding the implications of those changes in digital media involving audio is important work for the future. 

\section{Conclusion}
We began the process of quantifying how racial perceptions influence credibility in digital media (e.g., video conferencing, video posting, etc. ). This is a difficult problem to undertake as isolating the variables contributing to racial profiling is complicated. Here we isolated one of those variables -- visual representation (e.g., skin tone and facial features) -- and explore how modifying visual representation impacts credibility. We used AI to assist in creating and modifying specific race perceptions. CycleGAN was used to create unambiguous racial representations, while Deepfakes were used to incorporate subtle differences in modifying perceived race. We created surveys to test the image and video conditions to help understand these nuances. We used Amazon Mechanical Turk to recruit participants and obtain a more representative sample of the normal population compared to college students in a laboratory setting. All together, we crowd-sourced 800 participants to evaluate credibility (percent of participants who believed the speaker) in media shown in the surveys. By comparing the responses for the surveys associated with each condition, we measure the effect of how racial perceptions influence credibility. In the image condition, we found that the believed race of the speaker did not effect the credibility of that speaker. However, in the video condition where the racial adjustments are more nuanced, we found that the participants who believed the speaker was White were significantly more likely to believe that speaker was telling the truth (61.0\% versus 73.0\%). Additionally, we found that more positive sentiment was associated with the text responses for the altered video compared to the original video for participants justifying their credibility decisions. All together, our evidence suggests that subtle modifications to alter perceived race may allow unconscious bias to impact credibility. This has implications in the domains of sales, negotiations, job interviews and court trials when conducted in videoconferencing environments. We therefore recommend these systems implement full-disclosure and complete transparency when a person is altering their racial appearance using these platforms. We hope this work serves as a formal, initial investigation of this space and encourages further exploration. 


\bibliographystyle{aaai}
\bibliography{ref.bib}

\end{document}